
\magnification =\magstep1
\baselineskip = 15pt


\def\frac#1#2{{#1\over#2}}

\newcount\equationno      \equationno=0
\newtoks\chapterno \xdef\chapterno{}
\def\eqn{\eqno\eqname}
\def\eqname#1{\global \advance \equationno by 1 \relax
\xdef#1{{\noexpand{\rm}(\chapterno\number\equationno)}}#1}


\def\ga{\mathrel{\mathchoice {\vcenter{\offinterlineskip\halign{\hfil
$\displaystyle##$\hfil\cr>\cr\sim\cr}}}
{\vcenter{\offinterlineskip\halign{\hfil$\textstyle##$\hfil\cr>\cr\sim\cr}}}
{\vcenter{\offinterlineskip\halign{\hfil$\scriptstyle##$\hfil\cr>\cr\sim\cr}}}
{\vcenter{\offinterlineskip\halign{\hfil$\scriptscriptstyle##$\hfil\cr>\cr\sim\cr}}}}}


\def\n{\noindent}
\def\s{\smallskip}

\centerline{\bf NONLINEAR EVOLUTION OF DENSITY PERTURBATION USING}
\centerline{\bf APPROXIMATE CONSTANCY OF GRAVITATIONAL POTENTIAL}

\vfill

\centerline{\bf J.S. Bagla$^1$ and T. Padmanabhan$^2$}
\centerline{Inter-University Centre for Astronomy and Astrophysics}
\centerline{Post Bag 4, Ganeshkhind,}
\centerline{Pune - 411 007, INDIA.}

\vfill

\centerline{\bf ABSTRACT}
\medskip
\noindent
During the evolution of density inhomogeneties in an $\Omega=1$, matter
dominated universe, the typical density contrast changes from $\delta\simeq
10^{-4}$ to $\delta\simeq 10^2$. However, during the same time, the typical
value of the gravitational potential generated by the perturbations changes
only
by a factor of order unity. This significant fact can be exploited to provide
a new, powerful, approximation scheme for studying the formation of nonlinear
structures in the universe. This scheme, discussed in this paper, evolves the
initial perturbation using a Newtonian gravitational potential frozen in time.
We carry out this procedure for different intial spectra and compare the
results
with the Zeldovich approximation and the frozen flow approximation (proposed by
Mattarrese et al. recently). Our results are in far better agreement with the
N-body simulations than the Zeldovich approximation. It also provides a
dynamical explanation for the fact that pancakes remain thin during the
evolution. While there is some superficial similarity between the frozen flow
results and ours, they differ considerably in the velocity information. Actual
shell crossing does occur in our approximation; also there is motion of
particles along the pancakes leading to further clumping. These features are
quite different from those in frozen flow model. We also discuss the evolution
of the two-point correlation function in various approximations.
\vfill

\centerline {IUCAA - 14/93; April 93; Submitted for publication}
\vfill
\hrule
\s
email : $^1$jasjeet @iucaa.ernet.in   \hfill  $^2$paddy@iucaa.ernet.in

\eject

It is generally believed that structures like galaxies etc. formed through
the growth of density perturbations via gravitational instability. To
understand the formation of these structures, it is necessary to evolve the
initial perturbations to highly non-linear regimes, which turns out to be a
technically formidable task. Usually one applies linear perturbation theory to
evolve the inhomogeneities when they are small, but relies on extensive N-body
simulations to model the non-linear regime. As a result, we have only a limited
knowledge
of the physics of the non-linear regime.

It would, therefore, be worthwhile to develop semianalytic approximations which
could help us to understand the non-linear evolution of the perturbations. In
this paper, we propose one such approximation scheme. We describe qualitatively
its main features and compare its results with some other approximation
schemes.

To motivate the approximation scheme suggested in this paper, it is best to
begin by recalling some well known theoretical facts. Consider the evolution of
a density perturbation $\delta$ in the matter dominated era of an $\Omega=1$
Friedmann model, described by an expansion factor $a(t)$. It is well known that
$\delta \propto a$ for the growing mode, implying that the perturbed Newtonian
gravitational potential $\phi$ (generated by this perturbation) remains
constant in time. As evolution proceeds, $\delta$ will approach unity
invalidating the application of linear theory. Even though non-linear evolution
is extremely complex, it is reasonable to assume that the constancy of
gravitational potential is approximately maintained even in the non-linear
regime. For example, if one uses the spherical model (Peebles, 1980) to study
the non-linear evolution then the gravitational potential changes only by a
factor 2 or so as
 the structures turn around, collapse and virialise. After the bound structures
are formed, virial theorem ensures that no significant evolution occurs to the
 potential. {\it In other words, the gravitational potential due to
perturbations varies only by a factor of the order unity when density contrast
 changes from, say,  $10^{-4}$ to $10^2$}. We believe this is an extremely
important fact which can be profitably exploited to study non-linear evolution.

In order to implement this feature effectively, one can take a cue from another
popular approximation scheme, usually called the Zeldovich approximation
 (Zeldovich, 1970; Shandarin and Zeldovich, 1989). In the Zeldovich
approximation, one moves the particles using a fixed velocity field which is
 taken to be the initial velocity field. Mathematically, this scheme is
implemented by assuming that
$$ {\bf x} (t) = {\bf q} + a {\bf v} ({\bf q}) \eqn\qq $$
where ${\bf q}$ is the Lagrangian coordinate of a particle, ${\bf x}$ is the
co-moving Eulerian coordinate and ${\bf v}({\bf q}) = d{\bf q}/da$ is  the
 initial velocity field. This field ${\bf v}({\bf q})$ is related to the
gravitational potential $\phi$ by the relation
$$ {\bf v} \equiv \nabla\Phi \equiv  - \frac{2}{3a{\dot a}^2} \nabla \phi
\eqn\vp $$
Since Zeldovich approximation perturbs the trajectories rather than the
density,
 it can be used to study a larger range of density contrast than the linear
perturbation theory. Unfortunately this approximation suffers from two
 significant shortcomings. The density contrast at any time, calculated from
mass conservation, can be expressed as
$$ \delta = \left[ \prod\limits_{i=1}^3 (1 - a \lambda_i) \right]^{-1} -1
\eqn\qq $$
where $\lambda_i$ are the eigen values of the matrix ($- \partial {\rm v}_i /
\partial q_j$) with $\lambda_1 \ge \lambda_2 \ge \lambda_3 $. Hence, strictly
 speaking, the approximation breaks down when $a \lambda_1 = 1$ which occurs at
fairly low density contrasts (say, around $\delta \simeq 3$) in realistic
models.

A more serious shortcoming is the following: Zeldovich evolution gives a
distorted view of the density distribution once pancakes form and
particles move {\it through} the pancakes. Since the particles only `remember'
the {\it initial} velocities, they move past the pancakes unhindered, thereby
 leading to fair amount of thickening of pancakes. Numerical simulations, on
the
other hand, clearly show that pancakes remain thin for long periods of time.
We wish to emphasise that the question, ``why pancakes do not thicken during
the evolution of density perturbations ?'' is an important dynamical issue
which
needs to be understood from first principles. It is, of course, possible to
introduce artificial models, like the adhesion model containing damping
mechanisms, which correctly reproduce the N-body results. The key question, in
our opinion, is to understand why models like the adhesion model work as well
as they do. In other words, there is something intrinsic in the gravitational
dynamics of particles in an expanding universe which dampens the kinetic energy
in the component pependicular to the pancake more than the kinetic energy
parallel to the pancake. This effect can be understood qualitatively
 as follows :

Consider a set of particles interacting via Newtonian gravity in an
expanding universe. If we confine our attention to regions with
dimensions much smaller than the Hubble radius, then the equation of
motion for the $i^{th}$ particle is well approximated by:

$$
\ddot{\bf r}_i=- \sum_{j\neq i} \frac{Gm}{\mid{\bf r}_{ij}\mid^3}\,;\,\,\,{\bf
r}_{ij}={\bf r}_i-{\bf r}_j
\eqn\qq$$
Here ${\bf r}_i$ stands for the proper coordinate related to the comoving
coordinate ${\bf x}_i$ by ${\bf r}_i=a(t){\bf x}_i$.In terms of ${\bf x}_i$, we
have
$$
\ddot{\bf x}_i+\frac{2\dot{a}}{a}{\bf \dot{x}}_i=-\nabla\phi ;
\eqn\qq$$
where
$$
\nabla^2\phi = \frac{4\pi G}{a^3}\left[\sum_{j}m\delta({\bf x}-{\bf
x}_j)-\rho_0a^{3}_{0}\right]=4\pi G\rho_b\delta
\eqn\qq$$
in the matter dominated phase of the Friedmann model. In the above form the
equations depend explicitly on $t$ due to the presence of the terms
$(\dot{a}/a)$ and $a^3$. This can be avoided by introducing new dimensionless
time and space coordinates $\tau$ and ${\bf y}_i$ via
$$
\tau \equiv \ln(t/T);\,\,\,{\bf x}_i\equiv L{\bf
y}_i;\,\,\,L^3=Gmt^{2}_{0}/a^{3}_{0}
\eqn\qq$$
with an arbitrary constant $T$. Transforming the equations, we easily find
that:
$$
\frac{d^2{\bf y}_i}{d\tau^2}+\frac{1}{3}\frac{d{\bf y}_i}{d\tau}=-\nabla_yU
\eqn\dyn$$
$$
\nabla^{2}_{y}U=4\pi \sum_{i}\delta({\bf y}-{\bf y}_j)-\frac{2}{3}.
\eqn\qq$$
Equation \dyn\ can be used to understand the motion of the particles near a
pancake. Locally, the gravitational acceleration, ${\bf g}=-\nabla U$
is towards the pancake and has no appreciable component parallel to the
pancake.
Hence the velocity parallel to pancake (${\rm v}_\|)$ decays only due to the
cosmic
expansion (described by the $(1/3)\dot{\bf y}$ term) while the velocity normal
to the pancake $({\rm v}_\bot)$ is affected by the potential as well. It turns
out that this effect restricts the motion in direction perpendicular to the
pancake(see Fig.4 and our discussion later).
Quite clearly, the local gravitational force produced by the pancake plays an
important role in this process. Since Zeldovich approximation misses out the
 effects of acceleration and evolves particles using the initial velocity
field, it fails to reproduce the thin nature of the pancakes.

The above discussion suggests a possible new way of approximating non-linear
evolution. The success of Zeldovich approximation clearly demonstrates that one
 should work in a Lagrangian description and consider perturbation of
trajectories. However, it is also necessary that the effect of acceleration is
incorporated into this scheme. On a rigorous N-body simulation, this is done by
 using the {\it exact} gravitational force acting on the particle at every
instant of time. However, if our hypothesis of approximate constancy of
gravitational potential is correct, then one should be able to make
considerable
progress by evolving the trajectories using a prespecified  gravitaitional
potential which remains frozen in time. This is the idea which we pursue here.

The Euler equation governing the motion of fluid particles can be written in
the comoving coordinates as
$${d{\bf v}\over{da}} = -{3\over{2a}}\biggl[ \mu \nabla\phi + {\bf v}\biggr]
\eqn\eul$$
where the velocity is defined as ${\bf v}=d{\bf x}/da$ and $\mu$ is a constant
that equals $(2/3a{\dot a}^2)$. The left hand side of \eul\ is, of course, the
 convective derivative evaluated along the path of the particle.
In the proposed scheme we use the Euler's eqution with the potential as
specified at the initial time. The initial velocity of a particle is assumed to
 be $-\mu \nabla \phi$, just as in the Zeldovich approximation. At subsequent
instants of time, the accleration is computed by using the {\it instantaneous}
velocity of the particle and the {\it initially specified} potential at its
{\it
 current } position.  In reality, the potential will also change
with time as the density distribution evolves. It is this fact which we ignore
by invoking the hypothesis of constancy of the potential.

We find that this approximation works well and that the pancakes do not
thicken. The particles tend to move along the pancakes towards regions of lower
potential
 to end up in a few clumps. The acceleration we use in this approximation is
largest in regions where the instantaneous velocity vector points along with
the gradient of potential, as happens for particles after they cross the
pancake.

This may be thought of as a ``frozen potential'' formulation for the evolution
of density perturbations. Recently, Mattarrese et. al.(1992) have proposed a
 frozen flow approximation which essentially freezes the velocity field rather
than the potential. In this approach the particles move along streamlines
computed from the initial potential, using the same relation as in Zeldovich
 approximation \vp\ . However, here the inertia of particles is ignored,
whereas in Zeldovich approximation inertia is assumed to dominate over change
in the force field. We take into consideration both factors but assume a
constant
 potential. As we shall see,  our approximation works  better than the frozen
flow approximation.

The idea was explored numerically in the following manner. We start with
an initial potential $\phi$ (in 2D with a $64\times 64$ grid) which is a
 realisation of a gaussian random field with the power spectrum $P_\phi
\propto k^{(n-2)}$ for various values of $n$. (Note that $n$ in $2D$ is
``equivalent" to $n-1$ in $3D$). The particles are now moved by the ansatz
described above. We normalise our expansion factor such that $a=<\delta_L>$,
the average density contrast calculated by the linear theory. Thus we would
expect
nonlinear structures to form around $a\ga 1$.

In figures 1 and 2 we have shown the evolution for $n=1$ and $n=-1$
(corresponding to $n_{3D}=0,-2$). The top three frames are based on our
approximation, the middle three are based on frozen flow and the bottom
ones are from using the Zeldovich approximation. The time evolution proceeds
from left to right in all the cases and the frames correspond to $a=1,2$ and
$3$.

It is quite obvious that the Zeldovich approximation is the worst of the three
for $a\ga 1$. The pancakes thicken enoromously and -- of course -- this is to
be expected. The top and middle frames show that there is some supeficial
similarity in the spatial distribution calculated by frozen flow and frozen
potential; however, there is significant difference in the details. The
velocities of particles  in the nonlinear regimes (and the nature of their
motion near pancakes) are quite different in the two cases. In frozen flow, no
shell crossing occurs and the particles approach the pancakes more and more
slowly. In the frozen potential approximation, there is  shell crossing
and significant motion near the pancakes. To see this difference clearly,
we have plotted the trajectories of a few particles near the pancakes
in fig. 4. In the frozen flow, trajectories asymptotically approach one another
defining the ``pancakes" (thin lines in fig.4). But in the frozen potential
approximation (thick lines), the particles cross the pancake and oscillate
around the plane because of the local acceleration. It is this feature which
limits the thickness of the pancake. The pancake in figure 4 is shown at
$<\delta_L>=2$ while the trajectories are evolved for a much longer time to
show that particles remain confined in the pancake. One can also see that the
results of our approximation are similar to that of N-body simulations and that
in evolving from $a=2$
to $a=3$, the particles form more clumpy regions in comparison with frozen flow
approximation.

Figure 3 shows the frozen potential (top) and frozen flow (bottom) results for
$n=0$ (left) and $n=2$ (right). All frames are evaluated at $a=3$. The frozen
flow does not bring out the clumpiness expected in the $n=2$ case as well as
the frozen potential does. Comparison with figures 1,2 show that the
approximation works quite well for a wide variety of $n$ which are of interest.

The nature of these approximations -- Zeldovich, Frozen flow and Frozen
potential -- can be compared qualitatively in the following simple manner. Let
${\bf v} ({\bf q})$ be the initial velocity field which we start with as a
realization of some given power specturm. Suppose that, by random occurrence,
all the velocity vectors in some given, finite subregion of space point roughly
 towards a plane (on both sides). Quite clearly particles in that region will
move towards this plane and, in Zeldovich approximation, we will obtain a
 pancake coinciding with the plane. When we evolve the system further (using
Zeldovich approximation) the particles will continue to move with the {\it
original} velocity they have. This will result in the particles crossing  the
 plane and moving away from the plane and a consequent thickening of the
pancake.

In the frozen flow approximation the {\it local} velocity field is used to move
the particles (rather than the {\it initial} velocity field at the original
 Lagrangian coordinate). For the velocity field which is generally pointing
towards a plane, continuity demands the velocities would get smaller and
smaller near the plane. This would mean that the velocity of particles normal
to the
 plane will become smaller and smaller as they approach the would-be pancake.
Quite obviously no thickening of pancake will take place, since no pancake has
really formed.

In the frozen potential formulation, the evolution is quite different. Along
with the initial velocity field, we also have an initial acceleration field
 ${\bf g}$, defined in that region. Since the density contrast was small at the
initial epoch, allowing the use of linear theory, we know that ${\bf
g}=-\mu\nabla\phi$ and ${\bf v}$ point in the same direction. In other words,
 the acceleration vector field also points towards the plane. If we evolve the
particles using the acceleration field, then the particles which cross over the
pancake get pushed back towards the pancakes. This force prevents the particles
 from moving too far away from the pancake; consequently the pancakes remain
thin.

We stress that in the frozen flow approximation no actual shell crossing can
occur. Thus even though frozen flow approximation prevents thickening of
 pancakes, it does it in a somewhat crude manner. In the case of frozen
potential approximation, shell crossing does occur but the particles are
dynamically influenced by the gravitational potential which reverses their
 direction of motion when they cross the pancake.

In Fig.5 and 6 we have plotted the two point correlation function, as it
evolves with time, for the frozen potential and frozen flow approximation. The
distance
 scale used has been left arbitrary as we are working in two dimensions and
comparison with the real universe is not very meaningful. A somewhat
qualitative comparison can be made using the fact that the correlation function
is about
 unity at a grid size of $1.5$ at $a=1$ for frozen potential approximation. It
is clearly seen that for frozen flow approximation, the two point correlation
function is much more strongly peaked at small scales than the frozen potential
 approximation at $a=3$.

It would be interesting to compare our results with actual N-body simulations
in detail. We have to be content in this paper with visual comparisons with
published results since we do not possess an N-body code now. We hope to do
this
in a future colloboration.

There are several further directions and possibilities which we plan to
explore using this approximation. Since the velocity information in our
approximation is quite different from that of frozen flow, it is worthwhile
to calculate the scaled average pair velocity, $h(a,x)=[-v_{pair}(a,x)/
\dot{a}x]$ and compare it with the mean square fluctuation in
mass, $\sigma^2(a,x)$, at the same epoch and scale. Theoretical reasoning
(see Nityananda and Padmanabhan, 1993; Padmanabhan, 1992) as well as results of
numerical simulations (Hamilton et al., 1991) suggest that: (i) The function
$h(a,x)$ depends on $a$ and $x$ only through $\sigma^2$; i.e., $h (a,x)=
h[\sigma^2(a,x)]$ and (ii) $h\propto \sigma^2$ for $\sigma^2\ll 1$, {\it
rises to a value greater than unity} and falls to unity for $\sigma^2\gg 1$.
This overshooting of hubble velocity in the formation of bound structures
is unlikely to occur in the froxen flow models but will happen in the present
approach.

The approximation suggested here could also replace the adhesion model in
some contexts. It may be recalled that the adhesion models (with an
artificial viscosity term) were originally
introduced to keep the pancakes thin and to provide information regarding the
locations of the pancakes. The approach described here keeps the pancakes
thin because of  natural, dynamical reasons eliminating the need for and
introduction of a viscosity term. The approach can also provide the location of
 pancakes. What is more, the code for the present approximation is both
conceptually and numerically simple to implement as compared to the adhesion
 code. The formation and properties of the voids, for example, can be studied
quite easily by this approach.

The approximation outlined in this paper is ideally suited to study another
important problem: the dynamics of baryons in the dark matter potential
wells. Right now, attacking this problem using N-body simulations is extremly
hard and time consuming. The approximation schemes like adhesion model will
not be useful in this context; the frozen-flow, on the other hand, will provide
 a distorted picture of the motion near pancakes. The frozen potential
approximation, with a better velocity information, may give us valuable
insights into the physical processes at work while economising greatly on the
computation
 involved.

Lastly, one would like to study the limitation on the validity of the
approximation by trying to understand accuracy to which the gravitational
potential remains constant in the course of real evolution. In an $\Omega=1$
matter dominated model the potential does not change during the linear regime.
When the nonlinear phase begins, if we approximate overdense regions by
suitably
averaged regions with spherical symmetry then the potential changes in the
nonlinear phase only by a factor of order unity. The situation
becomes more complex when we take into account the deviations from the
spherical symmetry. Even though the gravitational potential at any one
point is contributed by matter distribution all over, there is some amount
of cancellation between the contributions from overdense and underdense regions
 in an expanding universe. This suggests that even in a realistic case,
the change in the potential is only of order unity in the nonlinear phase.
We plan to address these and related questions in a future publication (Bagla
 and Padmanabhan, 1993).

The authors thank K.Subramanian for useful discussions. One of us (JSB) would
like to thank Varun Sahni for introducing him to quasilinear approximations.
JSB
 is being supported by the Junior Research Fellowship of C.S.I.R., India.

\vfill
\eject

\beginsection{\bf References}

\item{} Bagla, J.S. \& Padmanabhan, T. (1993), work in progress.

\item{} Hamilton, A.J.S. et. al. (1991), ApJ, {\bf 374}, L1--L4

\item{} Matarrese, S. et. al. (1992), MNRAS, {\bf 259}, 437--452

\item{} Nityananda, R. \& Padmanabhan, T. (1993), {Scaling Properties of
Gravitational Clustering in the Non-linear regime}, IUCAA --12/93, preprint.

\item{} Padmanabhan, T. (1992), Current Science, {\bf 63}, 379

\item{} Peebles, P.J.E. (1980), {\it Large Scale Structure of the Universe},
Princeton University Press.

\item{} Shandarin, S.F. \& Zeldovich, Ya.B. (1989), Rev.Mod.Phys. {\bf 61},
185--220.

\item{} Zeldovich, Ya.B. (1970), Astron. Astrophys. {\bf 5}, 84.

\vfill
\eject

\beginsection{\bf Figure Captions}

\n{\bf Fig.1 } Evolution of density perturbations for various approximations:
time increases from left to right. These frames correspond to $<\delta_L>=1,2$
and $3$ respectively. The top row of frames is for frozen potential
 approximation, the middle shows evolution according to the frozen flow
approximation and the bottom row is for Zeldovich approximation. Power spectrum
used here has the index $n=1$.
\s
\n{\bf Fig.2 } Same as figure 1 but for $n=-1$.
\s
\n{\bf Fig.3 } Evolved perturbations at $<\delta_L>=3$ for frozen potential
(top row) and frozen flow (bottom row) approximations. These are for $n=0$
(left column) and $n=2$ (right column) power spectra.
\s
\n{\bf Fig.4 } Trajectories of a few particles near a pancake. The pancake is
shown at $<\delta_L>=2$, the trajectories are evolved for a much longer time to
show that particles remain confined in the pancake. Thin lines correspond to
 trajectories in frozen flow approximation and thick lines for frozen potential
approximation.
\s
\n{\bf Fig.5 } Evolution of the correlation function in frozen potential
approximation. Note that $a=<\delta_L>$.
\s
\n{\bf Fig.6 } Same as figure 5 but for frozen flow approximation.

\end